\documentclass[12pt]{article}
\usepackage{a4wide}
\usepackage{graphicx}

\newcommand{\be}{\begin{equation}}
\newcommand{\ee}{\end{equation}}
\newcommand{\ba}{\begin{eqnarray}}
\newcommand{\ea}{\end{eqnarray}}
\newcommand{\dis}{\displaystyle}

\newcommand{\order}{{\cal O}}

\begin{document}
\begin{titlepage}
\begin{flushright}
CAFPE/70-06\\
LU TP 06-03\\
UG-FT/200-06\\
hep-ph/0601197\\
January 2006
\end{flushright}
\vspace{2cm}
\begin{center}

{\large\bf The $B_K$ Kaon Parameter in the Chiral Limit}
\\
\vfill
{\bf  Johan Bijnens$^{a)}$, Elvira G\'amiz$^{b)}$ and
Joaquim Prades$^{c)}$} \\[0.5cm]

$^{a)}$  Department of Theoretical Physics, Lund University\\
S\"olvegatan 14A, S-22362 Lund, Sweden.\\[0.5cm]

$^{b)}$ Department of Physics \& Astronomy,
University of Glasgow\\ Glasgow G12 8QQ, United Kingdom.\\[0.5cm]

$^{c)}$ Centro Andaluz de F\'{\i}sica de las Part\'{\i}culas
Elementales (CAFPE) and Departamento de
 F\'{\i}sica Te\'orica y del Cosmos, Universidad de Granada \\
Campus de Fuente Nueva, E-18002 Granada, Spain.\\[0.5cm]

\end{center}
\vfill
\begin{abstract}
\noindent
We introduce four-point functions in the hadronic ladder resummation
approach to large $N_c$ QCD Green functions.
We determine the relevant one to calculate the $B_K$ kaon parameter
in the chiral limit.  This four-point function contains both 
the large momenta QCD OPE and the small momenta ChPT at NLO  
limits, analytically. 
We get $\hat B_K^\chi = 0.38 \pm 0.15$. 
We also give  the ChPT result at NLO for the relevant four-point function
to calculate $B_K$ outside the chiral limit, while the leading QCD OPE
is the same as the chiral limit one. 
\end{abstract}
\vfill
\end{titlepage}

\section{Introduction}

 Indirect kaon CP-violation  in the Standard Model
(SM) is  proportional to the matrix element
\ba
\label{matrix}
\langle \overline K^0 | K^0 \rangle
&=& -i C_{\Delta S=2} \, C(\nu) \, \langle \overline{K^0}|
\int {\rm d}^4 y \, Q_{\Delta S=2} (y) | K^0 \rangle \nonumber \\
&\equiv&-i C_{\Delta S=2} \, \frac{16}{3} \, \hat B_K \, F_K^2 m_K^2
\ea
with
\ba
\label{defQ}
Q_{\Delta S=2}(x) \equiv 4 L^\mu(x) L_\mu(x) \, ;
\hspace*{0.5cm} 2 L_\mu(x) \equiv
\left[\overline s \gamma_\mu (1-\gamma_5) d\right](x) \, .
\ea
and $C(\nu)$ a Wilson coefficient.
$C(\nu)$ and $C_{\Delta S=2}$ are known in
perturbative QCD to next-to-leading order (NLO).
A review can be found in \cite{rev}. 

The kaon bag parameter, $\hat B_K$, defined 
in (\ref{matrix})  is an important
input for the unitarity triangle analysis 
and its calculation has been addressed  many times in the past.
There have been four main QCD-based techniques used to calculate it:
QCD-Hadronic Duality \cite{PR85,PRA91},
three-point function QCD Sum Rules \cite{threepoint},
lattice QCD and the $1/N_c$ ($N_c =$ number of colors)
expansion. 
Recent reviews of the unitarity
triangle, where the relevant references for the inputs
can be found, are~\cite{BAT03}.
For recent advances using lattice QCD 
see~\cite{lattice04a,lattice04b,Dawson05}.

Here, we present a 
determination of the $\hat B_K$ parameter at NLO  in the
$1/N_c$ expansion. 
That the $1/N_c$ expansion introduced in~\cite{THO74,WIT79}
would be useful in this 
regard was first suggested by Bardeen, Buras and G\'erard
\cite{BBG} and reviewed by Bardeen in~\cite{BAR89,BAR99}. 
There one can find most of the references to previous work
and applications of this non-perturbative technique.

Work directly related to this work can be found in 
\cite{BP95,BP00} where a NLO in  
$1/N_c$ calculation of $\hat B_K$ within and outside
the chiral limit was presented. There, the relevant spectral function 
is calculated at very low energies using Chiral Perturbation Theory (ChPT),
at intermediate energies with the extended Nambu-Jona-Lasinio (ENJL) model~\cite{ENJL} 
and at very large energies with the operator product expansion (OPE).
The method, including how to deal with the short-distance scheme dependence,
has been discussed extensively in~\cite{BP00}. The improvement in the
present work is basically in the intermediate energies and the numerical
quality of the matching to short distances.

Another calculation of $\hat B_K$  in the chiral limit
at NLO in the $1/N_c$ expansion is in \cite{PR00}. There, 
the relevant spectral function is saturated by the pion
pole and the  first rho meson resonance --minimal hadronic 
approximation (MHA). In \cite{CP03}, the same technique as 
in \cite{PR00} was used  but including  also 
the effects of dimension eight operators in the OPE
of the $\Delta S=2$ Green's function and adding  the first
scalar meson resonance to the relevant spectral function.
This works differs from that in two aspects. We use our $X$-boson method
which allows for an easy identification of the precise scheme used in the
hadronic picture by only using currents and densities directly in four
dimensions rather than using dimensional regularization throughout.
This is a difference in formalism, not in physical content. On the other
hand, we use a much more elaborate way to include hadronic states allowing
the use of more constraints \cite{BGLP03} 
and thus more resonances than the work of~\cite{PR00,CP03}. 
The hadronic model developed in \cite{BGLP03} also 
includes current quark masses effects and  will be used 
to determine $\hat B_K$ in the real case \cite{BGP06a}.

The present manuscript is organized as follows. In Section~\ref{QCD}
we collect a few perturbative formulas at NLO in QCD for completeness.
Section~\ref{Xboson} explains in detail how we derive the value of
$\hat B_K$ from a large $N_c$ Green's function with two densities
and two currents. This can be found essentially in~\cite{BP00} but we
add here an extra QCD calculation of the short-distance constraint
analogous to the equivalent terms in our work 
on $Q_7$ and $Q_8$ \cite{Q7Q8}.
QCD is still not solved even at leading order
in the $1/N_c$ expansion. Therefore, at present, any NLO in $1/N_c$
approach to weak matrix elements needs a large $N_c$ hadronic model. 
We present
in Section~\ref{laddermodel} our hadronic model based on a severe
approximation to a ladder resummation of QCD as explained in~\cite{BGLP03}.
This model
is used at intermediate distances and is the main uncertainty in our
approach.
Then we discuss the chiral limit results in Section~\ref{BKchiral}.
The extension beyond the chiral limit needs more work both at 
 intermediate distances but we include some comments already
in Section~\ref{nonchiral}.  At short distances, the leading 
dimension six OPE operator is known and agrees with the chiral one. 
The last section contains our conclusions and a comparison
with other approaches.

Some preliminary results were already presented in \cite{BP03}
and \cite{PBG05}. 

\section{Some perturbative QCD formulas}
\label{QCD}

This section is a compilation of the perturbative QCD coefficients 
at NLO from~\cite{rev} needed in our analysis of $\hat B_K$ and 
derived in~\cite{NLO}.

The coefficients $C(\nu)$ and $C_{\Delta S=2}$ are known
in perturbative QCD at next-to-leading  order (NLO)
in $a\equiv \alpha_S /\pi$ in two schemes \cite{rev,NLO},
the 
't Hooft-Veltman (HV) scheme ($\overline{\rm MS}$ subtraction
and non-anti-commuting $\gamma_5$ in $D\neq 4$)
and in the Naive Dimensional Regularization (NDR) scheme
($\overline{\rm MS}$ subtraction
and anti-commuting $\gamma_5$ in $D\neq 4$). 
$C_{\Delta S=2}$ collects
known functions of the integrated out heavy particle
masses and Cabibbo-Kobayashi-Maskawa matrix elements, it can be found
in~\cite{rev,NLO}. Its actual form
 is very important for CKM analysis but not needed here.

The Wilson coefficient $C(\nu)$ is 
\be
C(\nu)= \left( 1 + a(\nu) \left[\frac{\gamma_2}{\beta_1}-
\frac{\beta_2 \gamma_1}{\beta_1^2} \right]\right)\, 
[ \alpha_s(\nu) ]^{\gamma_1/\beta_1}\,.
\ee
$\gamma_1$ is the one-loop $\Delta S=2$ anomalous dimension
\be
\gamma_1 = \frac{3}{2}\left(1-\frac{1}{N_c}\right) = 1\,.
\ee
$\gamma_2$ is the two-loop $\Delta S=2$ anomalous dimension
\cite{NLO}
\ba
\gamma_2^{\rm NDR}&=&
-\frac{1}{32}\left(1-\frac{1}{N_c}\right)
\left[17 + \frac{4}{3} (3-n_f)
+ \frac{57}{N_c}\left(\frac{N_c^2}{9}-1\right)\right] = -\frac{17}{48}
\, , 
\nonumber \\ 
\gamma_2^{\rm HV}&=&
\gamma_2^{\rm NDR}-\frac{1}{2}\left(1-\frac{1}{N_c}\right) \beta_1  \,.
\ea
$\beta_1$ and $\beta_2$ are the first two coefficients
of the QCD beta function 
\ba
\nu \frac{{\rm d} a(\nu)}{{\rm d} \nu}
&=& 
 {\dis \sum_{k=1}} \beta_k \, a(\nu)^{k+1} \, , 
\nonumber \\
\beta_1&=&-\frac{1}{6} \left[11 N_c -2 n_f\right]= -\frac{9}{2}
\, , \nonumber \\
\beta_2&=&-\frac{1}{24} \left[34 N_c^2-13 n_f N_c + 
3\frac{n_f}{N_c} \right] = -8\, . 
\ea
The explicit numbers are for $N_c=n_f=3$.

\section{$X$-boson Method and Known Constraints}
\label{Technique}
\label{Xboson}

The $X$-boson method was explained in detail in~\cite{BP00}.
It takes the idea of~\cite{BBG} of reducing the four-quark operator
in~(\ref{defQ}) to products of currents and follows through the full
scheme and scale dependence. In \cite{BP00} it was explicitly showed
how short-distance
scale and scheme dependences can be taken into account analytically 
in the $1/N_c$ expansion.  Here, we only sketch the procedure
introducing the notation. 

The effective action ${\bf \Gamma}_{\Delta S=2}$, 
\be
{\bf \Gamma}_{\Delta S=2}
\equiv - C_{\Delta S=2} \, C(\nu)  \int {\rm d}^4 y \, Q_{\Delta S=2}(y)
+ {\rm h.c.} \,,
\ee
contains all the
short-distance physics of the SM.
We replace it by the exchange of a colorless
heavy $\Delta S=2$ $X$-boson with couplings
\be
\label{defX}
{\bf \Gamma}_{\rm LD}
\equiv 2 \, g_{\Delta S=2}(\mu_C,\cdots)\, 
  \int {\rm d}^4 y \, X^\mu(y) \, L_\mu(y) \, + {\rm h.c.}\, . 
\ee
The coupling $g_{\Delta S=2}(\mu_C,\cdots)$ is obtained~\cite{BP00}
with an analytical short-distance matching using
perturbative QCD.\footnote{At energies small compared
to the $W$-boson mass but where perturbative QCD is still valid.}
Afterwards we only need to identify the current $L_\mu$ in the hadronic
picture, not the four-quark operator $Q_{\Delta S=2}$.

The matching leads to
\ba
\frac{g_{\Delta S=2}^2(\mu_C,\cdots)}{M_X^2}
\equiv C_{\Delta S=2} \, C(\nu) \, \left[ 1 + a \left( \gamma_1 
\log\left(\frac{M_X}{\nu}\right) + \Delta r \right) \right] \, .
\ea
 The one-loop finite term  $\Delta r$ is scheme dependent
\be
\Delta r^{NDR} = -\frac{11}{8} \left(1-\frac{1}{N_c}\right) = -\frac{11}{12}
\, ; \quad
\Delta r^{HV} = -\frac{7}{8} \left(1-\frac{1}{N_c}\right) \,  
\ee
and makes the coupling $|g_{\Delta S=2}|$ scheme independent
to order $a^2$. This coupling is also independent of the scale $\nu$
to the same order.
Notice  that there is no dependence on 
the cut-off scale $\mu_C$ --this feature
is general of four-point functions  which are
product of conserved currents~\cite{BP00}.
This procedure thus includes the standard
leading and next-to-leading resummation to all orders of 
the large logs in $[\alpha_S \log(M_W/\nu)]^n$ and 
$\alpha_S \, [\alpha_S \log(M_W/\nu)]^n$ including the 
short-distance scheme dependence.

In order to get at the matrix-element (\ref{matrix}), we calculate
a two-point Green function in the presence of the weak effective action,
with pseudo-scalar densities carrying kaon quantum numbers.
After reducing the kaon two-quark densities, the two-point function
(\ref{twopoint}) provides the matrix element in (\ref{matrix}) via
standard LSZ reduction. 
The two-point function is evaluated using the matching with the $X$-boson
effective action. 
We thus want to calculate the two-point function
\cite{BP95,BP00}
\ba
\label{twopoint}
{\bf \Pi}_{\Delta S=2}(q^2)
= i {\dis \int} {\rm d}^4 \, e^{i q \cdot x}\, 
\langle 0 | T \left(P_{\overline{K}^0}^\dagger(0)
P_{K^0}(x) \, e^{i {\bf \Gamma}_{\rm LD}}
\right) | 0 \rangle\, 
\ea
which we  need to order $g_{\Delta S=2}^2$. After reducing, we obtain
\ba
\langle {\overline K}^0(q) | e^{i {\bf \Gamma}_{\Delta S=2}}
| K^0 (q) \rangle &=&
\langle {\overline K}^0(q) | e^{i {\bf \Gamma}_{LD}}
| K^0 (q) \rangle \equiv -i C_{\Delta S=2} \, 
\frac{16}{3} \, \hat B_K \, q^2 F_K^2  \nonumber \\
&=& \int \frac{{\rm d}^4 p_X}{(2\pi)^4} \, 
\frac{g^2_{\Delta S=2}}{2} \, \frac{i g_{\mu\nu}}{p_X^2-M_X^2} \,
{\bf \Pi}^{\mu\nu}(p_X^2,q^2) \,   
\ea
where $q^2$ is the external momentum carried by the kaons.
The basic object is the reduced four-point function
\be
\label{basic}
{\bf \Pi}^{\mu\nu}(p_X^2,q^2) \equiv i^2 \, 4 \,
\langle \overline{K}^0(q) | \int {\rm d}^4 x \, \int {\rm d}^4 y \, 
e^{-i p_X \cdot (x-y)} \, T\left( L^\mu(x) \, L^\nu(y)
\right) | K^0(q) \rangle
\, .
\ee
This can be obtained from the four-point Green's function with
two kaon pseudo-scalar densities and two currents $L_\mu$.

At large $N_c$, the reduced four-point function (\ref{basic})
factorizes
into two disconnected two-point functions at all orders in quark masses
and external momentum $q^2$.
This disconnected part is
\be
g_{\mu\nu} {\bf \Pi}^{\mu\nu}_{\rm disconn.}(p_X^2, q^2)
= (2\pi)^4 \delta^{(4)}(p_X) \, 8 \, q^2 \, F_K^2\,,
\ee
which leads to the well-known large-$N_c$
prediction\footnote{In the strict large $N_c$ limit we also have $C(\nu)=1$.}
\be
\hat B_K^{N_c}= \frac{3}{4} \, .
\ee

At next-to-leading order in the $1/N_c$ expansion, one has
\be
\label{NLO}
\hat B_K = \frac{3}{4} 
\frac{g^2_{\Delta S=2}}{M_X^2 \, C_{\Delta S=2}}
\left[1-\frac{1}{16\pi^2 F_K^2}
{\dis \int_0^\infty} {\rm d} Q^2 \, F[Q^2] \right]
\ee
with $Q^2$ the X-boson momentum in Euclidean space and
\ba
\label{FF}
F[Q^2]&\equiv&
-\frac{1}{8\pi^2} 
\, {\dis \lim_{q^2\to m_K^2}}
\int {\rm d} \Omega_Q \, \frac{Q^2}{1+(Q^2/M_X^2)}
\, \frac{
  g_{\mu\nu}{\bf \Pi}^{\mu\nu}_{\rm conn.} (Q^2, q^2)}{q^2} \, .
\ea
The next point is the calculation of (\ref{FF}).
There are two energy regimes where we know
how to calculate  $F(Q^2)$  within QCD, namely, at
very large $Q^2$ and at very small $Q^2$.

In the first regime 
$Q^2 >> 1 \, {\rm GeV}^2$, with $q^2$ kept small and 
we can use the operator product expansion in QCD. One gets
\be
\label{OPEfour}
g_{\mu\nu} {\bf \Pi}^{\mu\nu}_{\rm conn.} (Q^2, q^2) = 
{\dis \sum_{n=2}^\infty}  \, 
{\dis \sum_{i=1}} \, \frac{C^{(i)}_{2n+2}
(\nu,Q^2) 
\langle {\overline K}^0 (q) | {\bf Q}^{(i)}_{2n+2} |
{K}^0 (q) \rangle}{Q^{2n}}\, ,
\ee
where ${\bf Q}^{(i)}_{2n+2}$  are local $\Delta S=2$
operators of dimension $2n+2$. In  particular,
\be
{\bf Q}_6 = 4 \, \int {\rm d}^4 x \, L^\mu(x) L_\mu(x)
\ee
 and 
\ba
\label{C6}
C_6(\nu,Q^2) = - 8 \pi^2 \gamma_1 a \,  
\left[1 + a \left[ \left(\beta_1-\gamma_1\right) \, 
\log \left( \frac{Q}{\nu} \right) + {\bf F}_1 \right] 
+ O(a^2) \right]
\ea
with\footnote{The leading terms differs from the one in~\cite{PR00} only
in terms subleading in $1/N_c$.}
\be
{\bf F}_1= \frac{\gamma_2}{\gamma_1}
+\left( \beta_1 -\gamma_1 \right)
\, \left[\frac{\Delta r}{\gamma_1} - \frac{1}{2} \right] = \frac{119}{16}\, 
\ee
where the term with ${\bf F}_1$ was not known before.
The finite term ${\bf F}_1$ is order $N_c$  and therefore this
 $a^2$ term is of the same order in $N_c$ as the leading term. In fact, 
at the
same order in $N_c$, there is an infinite series in powers of $a$.

The above can be used to take the limit $M_X\to\infty$ explicitly
via
\ba
\label{BKNLO}
\hat B_K &= &\frac{3}{4} 
C(\nu) \left(1+ a\left( \gamma_1 \, \log\left(\frac{\mu}{\nu}\right) 
+ \Delta r \right) \right)\times
\nonumber \\ &&
\Bigg[1-\frac{1}{16\pi^2 F_K^2}
\left({\dis \int_0^{\mu^2}} {\rm d} Q^2 \, F[Q^2]^{M_X\to\infty}
+\  {\dis \int_{\mu^2}^\infty} {\rm d} Q^2
F[Q^2]_{\rm D \geq 8}^{M_X\to \infty}\right)
\nonumber\\ &&
+ {\cal O} \left(\frac{\mu^2}{M_X^2} \right) +
{\cal O} (a^2) \Bigg] \, . 
\ea
Where $F[Q^2]_{\rm D \geq 8}$ is obtained inserting
 in (\ref{FF}) the result in (\ref{OPEfour}) 
 minus the dimension six term.

 For the list and a discussion of the dimension eight operators
see \cite{CP03,CDG00}. In \cite{CP03} there is a calculation
in the factorizable limit of the contribution of the dimension eight 
operators.  Numerically,  the finite term of order $a^2$ competes
with that contribution when $Q^2$ is around $(1\sim 2)$ GeV$^2$.

The second energy regime where we can calculate $F[Q^2]$
model independently is for $Q^2 \to 0$, where
the effective quantum field theory  of QCD is 
chiral perturbation theory. 
In ChPT the result is known up to order $p^4$ 
both in the chiral limit \cite{BP00,PR00} 
 and outside the chiral limit \cite{BP03}.
 The  result in the chiral limit is 
\be
\label{slope}
F^\chi[Q^2] = 3 + A_4 \, Q^2 + A_6\,Q^4+ \cdots\,,
\quad
A_4 =  -\frac{12}{F_0^2} \left(
2 L_1 + 5 L_2 + L_3 + L_9 \right)\,,
\ee
with $F_0$ the chiral limit of the pion decay constant
$F_\pi=92.4$ MeV. The next term, $A_6$, can be easily calculated 
but contains some of the 
unknown $C_i$ constants from the $p^6$ ChPT Lagrangian.

We still need to describe the intermediate energy region
for which we use the large $N_c$ hadronic model described 
in the next section. We have used that model
to predict the  series in (\ref{slope}) minus $3+ A_4 Q^2$, 
which is known. This is equivalent
to predict the relevant $C_i$ and higher couplings combinations.

\section{A Ladder Resummation Large $N_c$ Model}
\label{laddermodel}

All the results and information we have presented so far
on $\hat B_K$ are model independent.
In particular, we have seen in the previous section that there
are two energy regimes which can be calculated within QCD.
In this section, we describe a large $N_c$ hadronic model
that  provides the full ${\bf \Pi}^{\mu\nu}_{\rm conn.}(Q^2, q^2)$
which, apart from other QCD information,
contains these two QCD regimes analytically.

\begin{figure}[ht]
\centerline{\includegraphics[width=10cm]{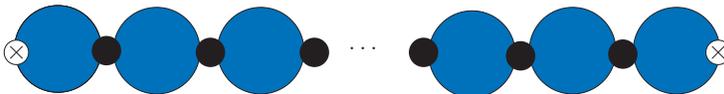}}   
\caption{Infinite geometrical series which gives 
full two-point functions at large $N_c$. The black vertices
that glue the vertex functions together are either $g_V$
for vector or axial-vector sources or $g_S$ for scalar and
pseudo-scalar sources.
 \label{fulltwopoint}}
\end{figure}

The large $N_c$ hadronic model we use was introduced in
\cite{BGLP03}. It can be thought of as QCD 
in the rainbow or ladder-resummation approximation.
The basic objects are vertex functions
with one, two, three, $\ldots$ two-quark
currents or density sources attached to them, referred to as one-point,
two-point, three-point, $\ldots$ vertex functions. These correspond to
the two-particle irreducible diagrams in large $N_c$ QCD.
\begin{figure}[ht]
\centerline{\includegraphics[width=10cm]{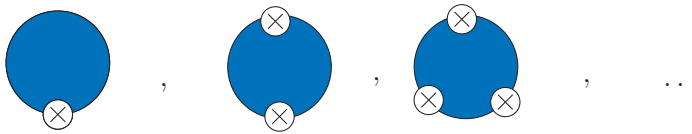}}   
\caption{ One-point, two-point, three-point, $\cdots$ vertex functions.
The crosses can be vector or axial-vector currents and scalar 
or pseudo-scalar densities.
 \label{vertex}}
\end{figure}
These vertex functions are glued into infinite
geometrical series with couplings $g_V$ for vector or
axial-vector sources and $g_S$ for scalar or pseudo-scalar
sources, what can be seen as a very crude approximation to the 
two-particle reducible part.
In this way one can construct full n-point
Green's functions in the presence of current quark masses
--see for instance, how to get full two-point functions
in Figure~\ref{fulltwopoint}.

The basic vertex functions in Figure~\ref{vertex}
have to be polynomials in momenta and quark masses to keep 
the large $N_c$ structure. As explained in~\cite{BGLP03},
only the first nonet of hadronic states  per channel, 
i.e., the pseudo-scalar pseudo-Goldstone bosons, the first vectors, 
the first axial-vectors and 
the first scalars can be easily generated within this framework.
The coefficients of the vertex functions are free constants
of order $N_c$, many of which can be fixed by imposing chiral 
Ward identities on the full Green's functions. 
Chiral perturbation theory at order $p^4$ and the operator
product expansion in QCD  help to determine
many more of these free coefficients in the vertex functions.
A consequence of the formalism described here
is that the vertex functions obey Ward 
identities with a constituent quark mass~\cite{BGLP03}.

The full two-point Green's functions obtained from the resummation
in Figure~\ref{fulltwopoint}
agree with the ones of large $N_c$ QCD when one limits the hadronic
content to be just one hadronic state per channel 
--our model does not produce any less or more constraints than
large $N_c$ QCD and all parameters can be fixed
in terms of resonance masses \cite{BP00} in agreement with
other groups using large $N_c$ and one state per channel.
Introducing two or more hadronic states per channel systematically
is difficult as explained in~\cite{BGLP03}.
We leave for future work the investigation into how to 
carry it out. 

Most low-energy hadronic effective actions used 
for large $N_c$ phenomenology are in the  approximation
of keeping the resonances below some hadronic scale and 
always in the chiral limit.
In many cases less than the four states included here are used.

The procedure described above of obtaining  full Green's functions
can be done in the presence of current quark masses.
In fact, in \cite{BP00}  two-point functions were calculated outside
the chiral limit  and all the new parameters that appear up
 to order $m_q^2$ can be determined except one;
namely,  the second derivative of the quark condensate with respect
to quark masses. Some predictions of the model we are discussing
involving coupling constants and masses
of vectors and axial-vectors in the presence of masses are
\ba
f^2_{Vij} \, M^2_{Vij} &=& f^2_{Vkl} \, M^2_{Vkl} \, , \nonumber \\
f^2_{Vij} \, M^4_{Vij} - f^2_{Vkl} \, M^4_{Vkl} &=& 
-\frac{1}{2} \langle \overline q q \rangle_\chi \, 
\left(m_i+m_j-m_k-m_l\right)\, \nonumber \\
f^2_{Aij} \, M^2_{Aij} + f^2_{ij}&=& f^2_{Akl} \, M^2_{Akl} 
+ f^2_{kl}\, , \nonumber \\
f^2_{Aij} \, M^4_{Aij} - f^2_{Akl} \, M^4_{Akl} &=& 
\frac{1}{2} \langle \overline q q \rangle_\chi \, 
\left(m_i+m_j-m_k-m_l\right)\, , 
\ea
where $i,j,k,l$ are indices for the up, down
and strange quark flavors.

\begin{figure}[ht]
\centerline{\includegraphics[angle=90,width=8cm]{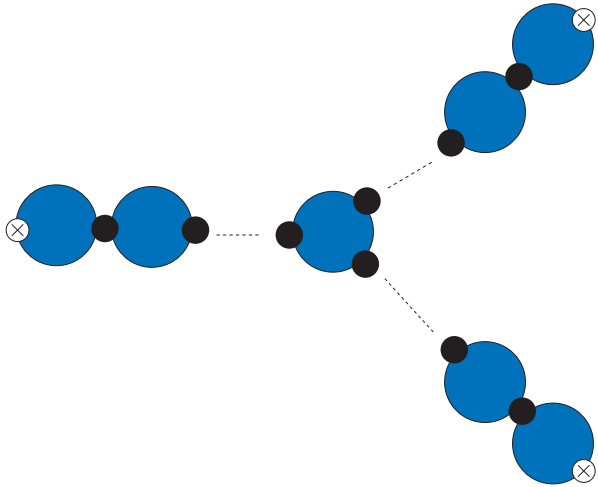}}   
\caption{ Infinite geometrical series which gives 
full three-point functions at large $N_c$. The crosses
that glue the vertex functions are either $g_V$
for vector or axial-vector sources or $g_S$ for scalar and
pseudo-scalar sources.
 \label{fullthreepoint}}
\end{figure}

Three-point functions, as shown in Figure~\ref{fullthreepoint},
were calculated in the chiral limit in \cite{BGLP03}.
We have now all the needed ones in the study of $\hat B_K$ 
also outside the chiral limit and they will be presented elsewhere \cite{BGP06b}. 
Some three-point functions have also been
calculated in other large $N_c$ approaches and in the chiral limit 
\cite{MOU95,KN01,CIR04,RPP03}
--for instance, PVV, PVA, PPV and PSP three-point functions, where
P stands for pseudo-scalar, V for vector, A for axial-vector
and S for scalar sources.
They agree fully with the ones we get in our model
when restricted to just one hadronic state per channel.

Four-point functions are constructed analogously
with the two-topologies dictated by large $N_c$, one with two three-point
vertex functions and one with a four-point vertex function.
The final four-point functions fulfill the correct chiral Ward identities
--including current quark masses and we impose short-distance QCD
constraints as well as the ChPT results at NLO 
to determine the free parameters allowed by the chiral Ward identities.

\section{Chiral Limit Results}
\label{BKchiral}

We now calculate $F[Q^2]$ of Eq.~(\ref{FF}) in the chiral limit 
using the model of the previous section.
After integrating over the four-dimensional Euclidean
solid angle $\Omega_Q$ and doing the limit $q^2 \to 0$
as in (\ref{FF}), we get
\ba
\label{FFchi}
F^\chi[Q^2] &=& \delta_1 Q^2 + \delta_2
+\frac{\alpha_V}{Q^2+M_V^2} + \frac{\alpha_A}{Q^2+M_A^2}
+ \frac{\alpha_S}{Q^2+M_S^2}
+ \frac{\beta_V}{\left(Q^2+M_V^2\right)^2}
 \nonumber \\ 
&+& \frac{\beta_A}{\left(Q^2+M_A^2\right)^2}
+ \frac{\gamma_V}{\left(Q^2+M_V^2\right)^3} +
\frac{\gamma_A}{\left(Q^2+M_A^2\right)^3} \, .
\ea
We have used here the two-point vertex functions to second order 
in momenta,
higher orders will make it impossible to satisfy the Weinberg sum rules.
The three- and four-point vertex functions have been expanded to fourth order
in the external momenta. The short-distance constraints used are the
Weinberg sum rules and the fact that the vector form factor should vanish
as $1/Q^2$ for large $Q^2$. As expected from the arguments in \cite{BGLP03}, 
we find clashes between short-distance constraints  
for three-point functions and those of the needed four-point 
function here.  I.e., with just a finite number of hadronic states
 per channel  not all short-distance  constraints  for n-point Green
are compatible in general. 
Of course, one can always impose the latter which are  compatible
with the  subset of the short-distance constraints corresponding to
the momenta structure of  the three-point functions entering in the 
four-point functions but not with all three-point functions
short-distance constraints.    That is what 
we have done for calculating $\hat B_K$.

The short-distance constraints on $F[Q^2]$ discussed in Sect.~\ref{Xboson}
and Ref.~\cite{PR00,CP03} require $\delta_1=\delta_2=0$ in (\ref{FFchi}).
Imposing these we obtain the expression for $F[Q^2]$ given 
in the appendix.

The explicit calculation reveals that diagrams with the exchange
of vector and axial-vector states produce not only single poles
but also double and triple poles. These also arise from diagrams
with only one of these propagators due to the factor $1/q^2$
present in (\ref{FF}).\footnote{Ref.~\cite{PR00,CP03} use a different
underlying four-point function. The same pole structure does
show up there as well as is required by chiral symmetry.}

The function $F^\chi[Q^2]$ in (\ref{FF})
reproduces the large $N_c$-pole structure
found in \cite{PR00,CP03} but including 
the first hadronic state in all the spin zero and one channels.

The relevant free parameters are the masses
$M_V$, $M_A$ and $M_S$, the pion decay constant in the chiral limit, $F_0$,
and three combinations of constants appearing in the three- and four-point
vertex functions, $A_{1,2,3}^\chi$.

The low energy inputs we use are $F_0$ and the $\order(p^4)$ couplings 
$L_i$ determined from the ChPT fits to data.
We use two different fits, namely, fit 10 of~\cite{ABT01} for
$L_{1,2,3}$ and $L_{9,10}$ fit from \cite{BT02}, both with the full $p^6$
fits and the ones using only $p^4$ expressions.
The input values for these two cases $p^6$ and $p^4$ are given
together with some derived quantities in Table~\ref{tabLi}.
\begin{table}
\centerline{
\begin{tabular}{|c|c|c|}
\hline
 & $p^6$ input & $p^4$ input \\
\hline
$F_0$           & 87.7~MeV           & 81.1~MeV                 \\
$10^3\,L_1$     & 0.43               & 0.38                     \\
$10^3\,L_2$     & 0.73               & 1.59                     \\
$10^3\,L_3$     & $-$2.3             & $-$2.91                   \\
$10^3\,L_5$     & 0.97               & 1.46                     \\
$10^3\,L_9$     & 5.93               & 6.9                     \\
$10^3\,L_{10}$  &$-$4.4              &$-$5.5                    \\
$M_V$           & 805~MeV            & 690~MeV                  \\
$M_A$           & 1.16~GeV           & 895~MeV                 \\
$M_S$           & 1.41~GeV           & 1.06~GeV                 \\
$A_4$           & $-$12.7~$GeV^{-2}$ & $-$23~$GeV^{-2}$       \\
\hline
\end{tabular}
}
\caption{The low energy inputs used together with some derived quantities
for the two sets of input.}
\label{tabLi}
\end{table}
The values of the masses and the slope $A_4$
follow from the $L_i$ and $F_0$ used as input
using the relations of~\cite{BGLP03} and (\ref{slope}).
One could also use the physical masses and then use the $L_i$ only to get at
the slope. The case with $p^6$ input is within 30\% of the physical
masses. For the scalars, which mass to use is still an open question,
but with the mounting evidence that the first one are not present
in large $N_c$,~\cite{CENP03} and references therein, a mass of around
1.4~GeV seems fine.

We determine $A_1^\chi$ from the slope $A_4$ as given in (\ref{slope})
and the numerical values from Table~\ref{tabLi}.

$A_2^\chi$ and $A_3^\chi$ are obtained using the short distance constraints
of (\ref{OPEfour}), via
\be
\label{FFSD}
F[Q^2]_{SD}
 = \frac{D_6}{Q^2}+\frac{D_8}{Q^4}+{\cal O}\left(\frac{1}{Q^6}\right)\,.
\ee
We use the estimate 
$D_8/D_6\approx 0.13$~GeV$^{-2}$ of~\cite{CP03} but with rather large
uncertainties.
$D_6$ is calculated from (\ref{C6}) 
which contains the value of $\hat B_K$ inside it. One can choose here
the large $N_c$ values or the values which come out of the analysis.
We choose the latter ones using $\hat B_K=0.17,0.37$ for the $p^6,p^4$ 
inputs, respectively.
With $\alpha_S(m_\tau) = 0.35$ we obtain $D_6 = 0.0.028,0.052~$GeV$^2$
and $D_8 = 0.0036,0.0068~$GeV$^4$. In the remainder we use the 
short-distance (\ref{FFSD}) with these values. 

We have then used two different
ways to match the model $F[Q^2]$ to short distance. The first one is to
have the two coincide at the values of $Q^2= 2$ and 3~GeV$^2$,
these are labeled with ``values'' 
in the figures. The other is to have the model
reproduce the values of $D_6$ and $D_8$ in its large $Q^2$ expansion,
this case is labeled ``order'' in the figures.

\begin{figure}
\begin{minipage}{0.49\textwidth}
\includegraphics[width=\textwidth]{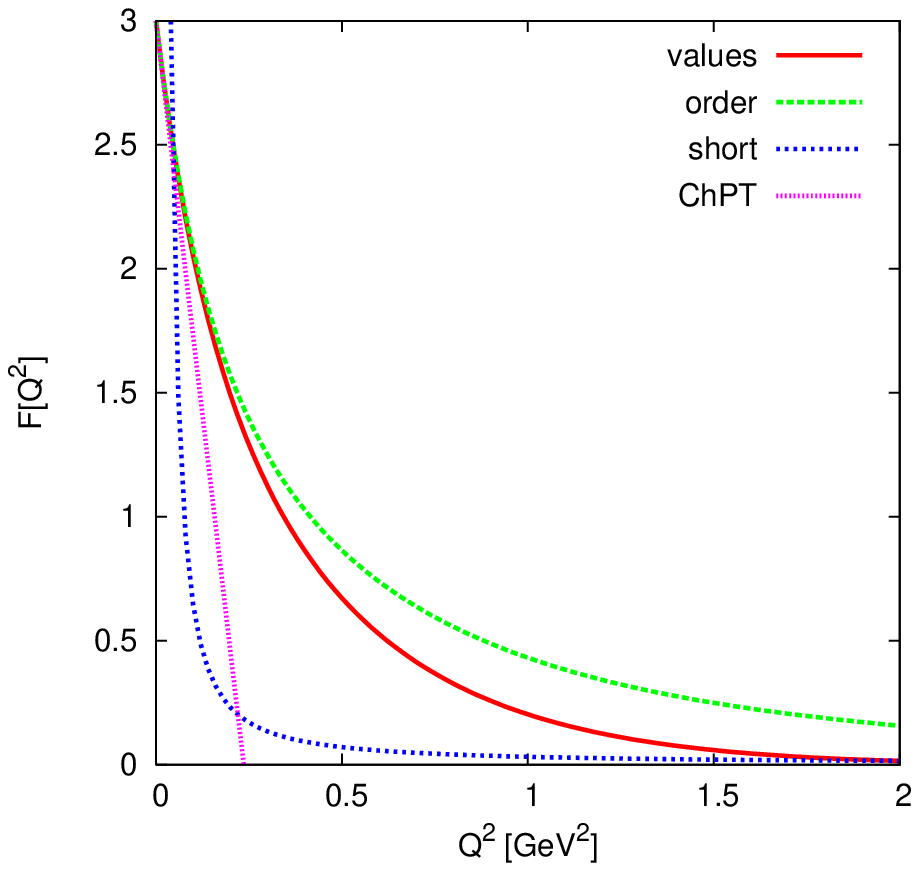}
\centerline{(a)}
\end{minipage}
\begin{minipage}{0.49\textwidth}
\includegraphics[width=\textwidth]{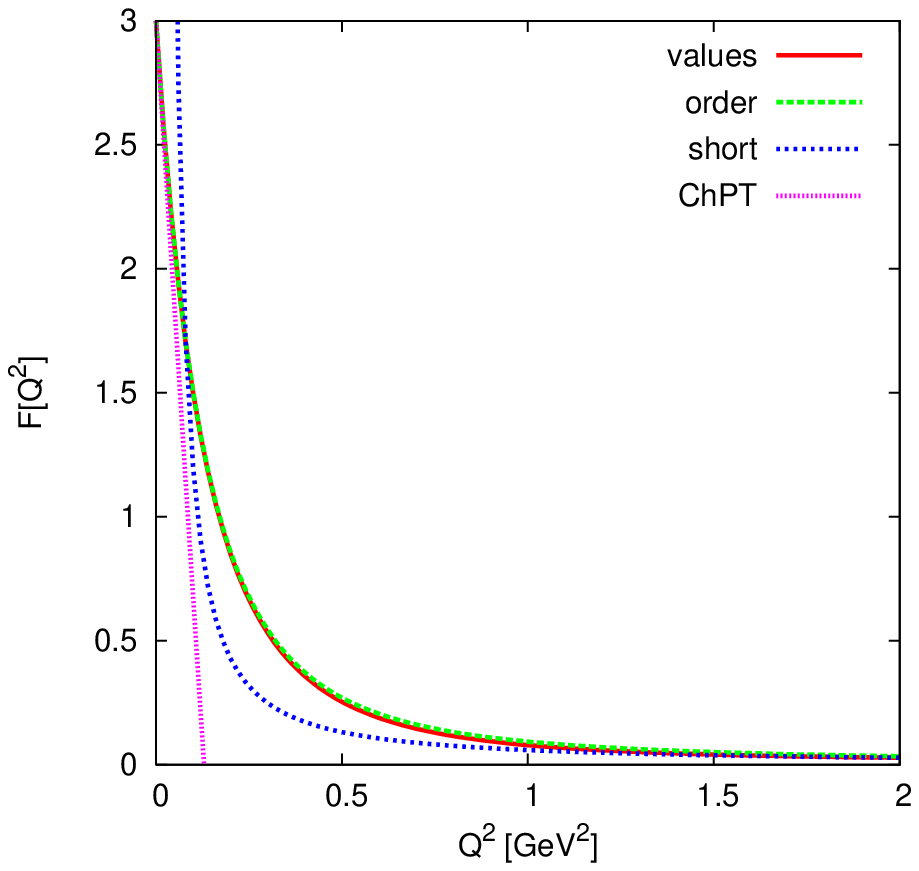}
\centerline{(b)}
\end{minipage}
\caption{$F[Q^2]$ from ChPT, short distance and the large $N_c$ ladder
resummation model interpolation. (a) with the $p^6$ fit values as input
(b) with the $p^4$ fit values as input.}
\label{figFQ2}
\end{figure}

In Fig.~\ref{figFQ2} we have plotted the short-distance curve and the model
curves as well as the order $p^4$ ChPT approximation.

\begin{figure}[t]
\centerline{\includegraphics[width=10cm]{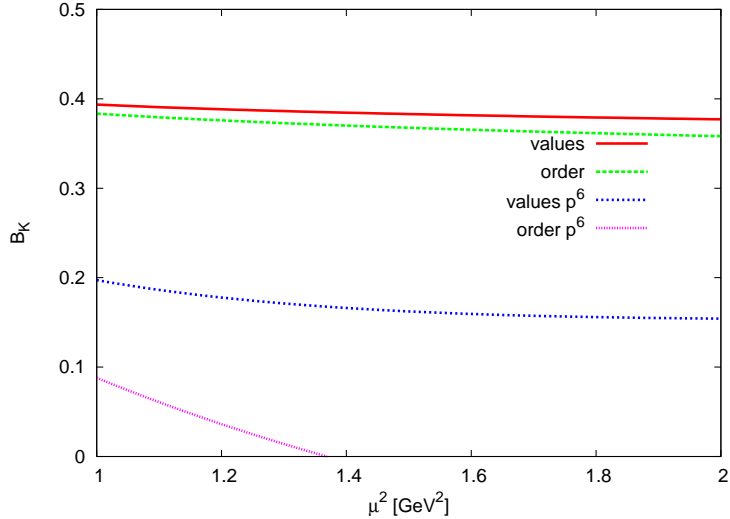}}   
\caption{$\hat B_K^\chi$ plotted vs the upper
limit of the integral in (\ref{FF}). See text for 
further explanation.
 \label{BKplot}}
\end{figure}

We now use (\ref{BKNLO})  and integrate 
(\ref{FF}) up to the matching point $\mu$ and from that
point on we use the OPE result including dimension
eight corrections \cite{CP03} of (\ref{FFSD}). 
This OPE contribution to 
$\hat B_K^\chi$ is
negligible. 
In Figure \ref{BKplot}, we plot this result as a function of $\mu^2$
Notice the nice plateau one gets between 1  and 2 GeV$^2$,
with the exception of the case ``order'' with the $p^6$ 
input. From Fig.\ref{figFQ2}a it can be seen that this because this case has
an extremely slow approach to the short-distance.
Varying the inputs within does not change the result more than
those shown. We thus obtain
\be
\hat B_K^\chi = 0.38 \pm 0.15 \, 
\ee
which is fully compatible with the one found
previously also in the chiral limit in \cite{BP95,BP00,PR00,CP03}.

\section{Outside the Chiral Limit Results}
\label{nonchiral}

The QCD results of the two energy regimes explained in 
Section \ref{Technique} 
are known outside the chiral limit too.

The function $F[Q^2]$ is known for very large values of $Q^2$,  
which can be calculated using the OPE in QCD.
In fact, the dimension six operator and Wilson coefficient
are the same as the chiral limit ones. Differences will first start
to appear at dimension 8 and we expect them to be small given the overall
small estimate of the dimension 8 corrections.

At small values of $Q^2$ we can use ChPT.
It is easy to get that $F[0]=0$ for the real case 
instead  of $F^\chi[0]=3$ in the chiral limit. Notice  
that the value of $F^\chi[0]$ is a strong constraint on  the value
of $\hat B_K^\chi$. That $F[0]=0$ outside the chiral limit is a strong
indication that corrections to the leading in $N_c$ value $B_K=3/4$
will be much smaller than in the chiral limit case.

The ChPT calculation in the real quark masses case
of $F[Q^2]$ has been done to order $p^4$ \cite{BP03} 
and  $F[Q^2]$ remains small  --below 0.15-- up to energies
around  0.2 GeV$^2$ and then goes negative.

Outside the chiral limit and to chiral order $p^2$, we get
\ba
F[Q^2]^{(2)} &=&
-\frac{Q^6+2 m_K^2 Q^4+ 2 m_K^4 Q^2}{m_K^2 (Q^2+m_K^2)^2}  +
 \frac{1}{4} \frac{Q^2+ 2 m_K^2 + 2 m_\pi^2}{m_K^2} \, 
I[Q^2,m_K^2,m_\pi^2]
\nonumber\\
&& +
\frac{3}{4} \frac{Q^2+ 2 m_K^2 + 2 m_{\eta_8}^2}{m_K^2} \, 
I[Q^2,m_K^2,m_{\eta_8}^2]\,.
\ea
To chiral order $p^4$ and leading in  $1/N_c$ we get
\ba
F[Q^2] &=& F[Q^2]^{(2)} - 4\,(2 L_1 + 5 L_2 + L_3) 
  \frac{ 3Q^6+6 m_K^2 Q^4+4 m_K^4 Q^2}{ F_\pi^2 (Q^2+m_K^2)^2} 
\nonumber\\&&
 + 8 L_5 \frac{Q^6+2 m_K^2 Q^4+2 m_K^4 Q^2}{F_\pi^2 (Q^2+m_K^2)^2} 
 + 16 L_8 \frac{m_K^2 Q^4}{F_\pi^2 (Q^2+m_K^2)^2} 
+ 4 L_9 \frac{Q^4}{m_K^2 F_\pi^2}
\nonumber\\&&
-  2 L_5 \frac{ Q^2 m_K^2-m_\pi^4+m_K^4+4 m_\pi^2 m_K^2}{m_K^2 F_\pi^2}
I[Q^2,m_K^2,m_\pi^2]
\nonumber\\&&
-  6 L_5 \frac{ Q^2 m_K^2-m_\eta^4+m_K^4+4 m_\eta^2 m_K^2}{m_K^2 F_\pi^2}
I[Q^2,m_K^2,m_{\eta_8}^2] 
\nonumber\\&&
-  L_9 \frac{Q^4+ (m_K^2-m_\pi^2)^2+ 2 Q^2 (m_K^2+m_\pi^2)}{m_K^2 F_\pi^2}
I[Q^2,m_K^2,m_\pi^2]
\nonumber\\&&
-  3 L_9 \frac{Q^4+ (m_K^2-m_{\eta_8}^2)^2+ 
2 Q^2 (m_K^2+m_{\eta_8}^2)}{m_K^2 F_\pi^2}
I[Q^2,m_K^2,m_{\eta_8}^2]\,.
\ea
The angular integration results in the function
\be
I[x,m_1,m_2] \equiv
\frac{x-m_1+m_2}{2 m_1} \left[\sqrt{1+4 m_1\,x/(x-m_1+m_2)^2}-1\right] \,.
\ee
We have used the Gell-Mann-Okubo relation to simplify the $p^4$ contribution
and dropped all terms that are subleading in $1/N_c$ as well as changed $F_0^2$
in the higher order to $F_\pi^2$.

\begin{figure}[t]
\centerline{\includegraphics[width=10cm]{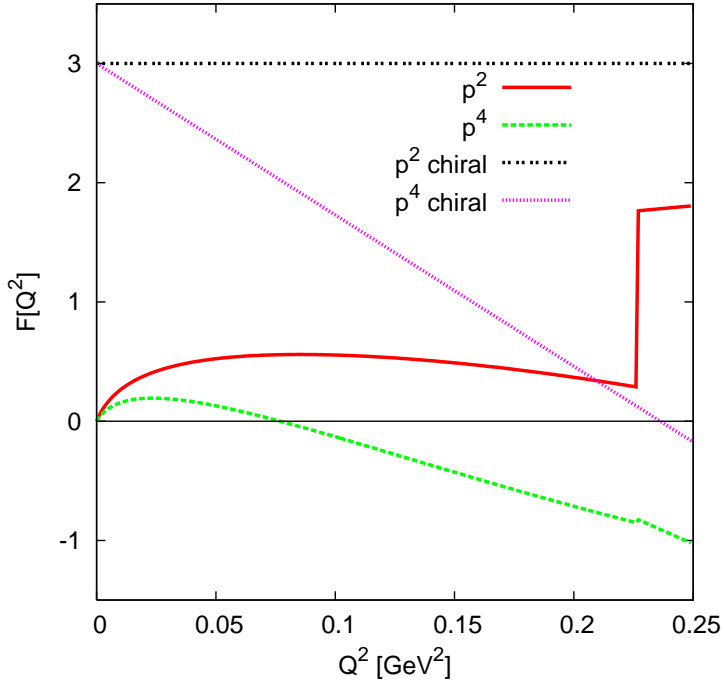}}   
\caption{The ChPT calculation of $F[Q^2]$ in and away from the chiral limit.}
 \label{figchpt}
\end{figure}

In Fig.~\ref{figchpt} we have shown $F[Q^2]$ at low $Q^2$ from the formulas
above in ChPT for definiteness with the $p^6$ inputs of Table \ref{tabLi}.
It can be seen that the function starts at zero and the area under the curves,
 which gives the corrections is much smaller away from the chiral limit.
The discontinuity in the curve is due to the fact that to the right
the state with an intermediate pion can go on-shell. 

One expects that higher ChPT order terms correct the behaviour
of $F[Q^2]$ for $Q^2>$ 0.2 GeV$^2$
and $F[Q^2]$ will  tend to the chiral limit curve
in Figure \ref{figFQ2}. The  question  is at which energy does it happens. 
This can only be answered using a hadronic large $N_c$
approximation to QCD at present. So, though we have strong indications
that the value $\hat B_K=3/4$ has small chiral corrections as
shown before, we have to wait till we get the full four-point
Green's function in the real case (\ref{basic}) to confirm it. 
The number of new free parameters in our hadronic model including quark
masses is rather high and we have at present not been able to determine all
as we did in the chiral limit discussed earlier.
We will eventually present the full result in the 
presence of current quark  masses in \cite{BGP06a}.

\section{Summary and Conclusions}
\label{conclusions}

 In the past few years,  lattice QCD  has produced many calculations
of $B_K$ using  different fermion formulations 
--see \cite{lattice04a,lattice04b,Dawson05} for references. 
Those include some chiral limit  
extrapolations --like for instance  in \cite{staggered} 
obtaining $\hat B_K^\chi=0.32\pm0.22$ with quenched staggered quarks,
 in \cite{AOK04}
obtaining $\hat B_K^\chi=0.34\pm0.02$ with two dynamical
domain wall quark flavours or in \cite{AOK05}
obtaining $\hat B_K^\chi=0.39\pm0.03$ with quenched domain wall quarks
--which show a clear decreasing tendency 
with respect to the real case value, in agreement with the 
results found here and in \cite{BP95,BP00,PR00,CP03}. 
Promising preliminary unquenched results have also started to appear 
in the last two years \cite{AOK04,latticeunquenched}.

The  QCD-Hadronic Duality result for $\hat B_K$\cite{PR85,PRA91} 
 is very close to the chiral limit result above because 
--as already mentioned in \cite{PRA91}-- what
was calculated there, is the order $p^2$ coefficient
of the chiral expansion which actually is the chiral limit value
of $\hat B_K$.

We have found values in this paper which are very compatible with the
those chiral limit values and in agreement with those of
\cite{BP95,BP00,PR00,CP03}. The improvement over our earlier work
\cite{BP95,BP00} is the use of a more reliable model for the intermediate
energy regime. What we have done beyond the work of \cite{PR00,CP03}
is to include more resonances and a somewhat different input 
for the low-energy
constants. In addition we have argued that away from the chiral limit the
corrections to the LO large $N_c$ value 3/4 should be much smaller.
A full result in the presence of current quark 
masses will be presented in \cite{BGP06a}.

\section*{Acknowledgments}

J.P. thanks the Department of Theoretical
Physics at Lund University where part of his work was done 
for the warm hospitality.
This work has been supported in part by
the European Commission (EC) RTN Network EURIDICE
Grant No.  HPRN-CT-2002-00311,
the EU-Research Infrastructure
Activity RII3-CT-2004-506078 (HadronPhysics),
the Swedish  Science Foundation,
MEC (Spain) and FEDER (EC) Grant No.
FPA2003-09298-C02-01, and by the Junta
de Andaluc\'{\i}a Grant No. FQM-101. 
E.G. is indebted to the EC for the
 Marie Curie Fellowship  No. MEIF-CT-2003-501309.

\appendix

\section{The Explicit Form of $F[Q^2]$}

The function $F[Q^2]$ has in the chiral limit
the form of Eq.~(\ref{FFchi}). The short distance already imposes
\be
\delta_1=\delta_2=0\,.
\ee
The remaining constants are
\ba
\alpha_V &=&  - \frac{6 M_V^6}{\left(M_V^2-M_A^2\right)^2}
              -\frac{3 M_V^4}{M_V^2-M_A^2}
              - \frac{3}{4} M_V^2
          + \frac{15}{4}\frac{M_V^4}{M_A^2}
         + \frac{12 M_V^8}{M_A^2} A_3^\chi
       +\frac{3}{128} M_V^6 N_\pi
\nonumber\\
\alpha_A &=&
           \frac{6 M_V^6}{\left(M_V^2-M_A^2\right)^2} 
          +\frac{3 M_V^4}{M_V^2-M_A^2}
          - \frac{3 M_V^4 M_S^2}{\left(M_A^2-M_S^2\right)^2}
          + \frac{15}{2}\frac{M_A^4}{M_V^2}
          - \frac{33}{4} M_V^2
          + \frac{9}{4} \frac{M_V^4}{M_A^2}
          - \frac{9}{2} M_A^2
\nonumber\\ &&
          + \frac{12 M_V^6}{M_A^2} A_1^\chi
          - \frac{12 M_S^6}{\left(M_A^2-M_S^2\right)^2} A_2^\chi
          + 12 M_S^2 A_2^\chi
          - 12 M_V^6 A_3^\chi
       + \frac{1}{384}\frac{M_A^8}{M_V^2} N_\pi
\nonumber\\
\alpha_S &=& 
  \frac{3 M_V^4 M_S^2 +12 M_S^6 A_2^\chi}{\left(M_A^2-M_S^2\right)^2}
\nonumber\\
\beta_V &=&
          - \frac{3 M_V^6 M_A^2}{\left(M_V^2-M_A^2\right)^2}
          + \frac{9}{4} M_V^4
          - \frac{3 M_V^6}{M_A^2}
          - \frac{6 M_V^{10}}{M_A^2} A_3^\chi
          -\frac{3}{64}M_V^8 N_\pi
\nonumber\\
\beta_A &=&
             \frac{3 M_V^8}{\left(M_V^2-M_A^2\right)^2}
          -  \frac{9 M_V^6}{M_V^2-M_A^2}
          - \frac{3 M_V^4 M_S^2}{M_A^2-M_S^2}
          - \frac{15}{2}\frac{M_A^6}{M_V^2}
          + \frac{3}{2} M_V^2 M_A^2
          + \frac{3}{4} M_V^4
          + 3 M_A^4
\nonumber\\&&
          - 12 M_V^6 A_1^\chi
          - \frac{12 M_A^4 M_S^2}{M_A^2-M_S^2} A_2^\chi
          + 6 M_V^6 M_A^2 A_3^\chi
       -\frac{1}{192}\frac{M_A^{10}}{M_V^2} N_\pi
\nonumber\\
\gamma_V &=&
       -\frac{3 M_V^6 M_A^2}{M_V^2-M_A^2}
       +\frac{3}{128} M_V^{10} N_\pi
\nonumber\\
\gamma_A &=&
       -\frac{3 M_V^2 M_A^6}{M_V^2-M_A^2}
       +\frac{1}{384}\frac{M_A^{12}}{M_V^2} N_\pi
\ea
The parts proportional to $N_\pi$ defined by
\be
N_\pi = \frac{N_c^2}{F_0^4\pi^4}\left(1-\frac{M_V^2}{M_A^2}\right)\,,
\ee
come from the diagrams with anomalous three-point vertex functions.


\begin{thebibliography}{0}

\bibitem{rev} 
A. J. Buras,
``Weak Hamiltonian, CP violation and rare decays,''
Les Houches lectures 1998, hep-ph/9806471;
G. Buchalla, A.J. Buras and M.E. Lautenbacher,
Rev.\ Mod.\ Phys.\  {\bf 68} (1996) 1125.


\bibitem{PR85} 
A. Pich and E. de Rafael, Phys. Lett. 
{\bf B 158} (1985) 477.

\bibitem{PRA91} 
J. Prades, C.A. Dom\'{\i}nguez,
J.A. Pe\~narrocha, A. Pich and E. de Rafael,
Z. Phys. {\bf C 51} (1991) 287.

\bibitem{threepoint} 
K.G. Chetyrkin et al. 
Phys. Lett. {\bf B 174} (1986) 104;
R. Decker, Nucl. Phys. {\bf B 277} (1986) 660;
N. Bili\'c, B. Guberina and C.A. Dom\'{\i}nguez,
Z. Phys. {\bf C 39} (1988) 351;
R. Decker,  Nucl. Phys. B (Proc. Suppl.) {\bf 7A} (1989) 180.

\bibitem{BAT03} 
``The CKM Matrix and the Unitarity
Triangle'', 
M. Battaglia, A.J. Buras, P. Gambino,
and A. Stocchi (eds), CERN (2003),  hep-ph/0304132.
J.~Charles {\it et al.}  [CKMfitter Group],
Eur.\ Phys.\ J.\ C {\bf 41} (2005) 1;
M.~Bona {\it et al.}  [UTfit Collaboration],
J. High Energy Phys. {\bf 07} (2005) 028.

  
\bibitem{lattice04a} 
M. Wingate, 
Nucl. Phys. B (Proc. Suppl.) {\bf 140} (2005) 68.

\bibitem{lattice04b}
S. Hashimoto in Proc. of ICHEP 2004, Vol I, p. 77,  
H. Chen {\it et al}, (eds),
World Scientific (2005), hep-ph/0411126.

\bibitem{Dawson05}
C.~Dawson, PoS {\bf LAT2005} (2005) 007.

\bibitem{THO74} 
G. 't Hooft, Nucl. Phys. { \bf B 72} (1974) 461;
bid. {\bf B 75}  (1974) 461.

\bibitem{WIT79} E. Witten, 
Nucl.\ Phys. {\bf B 160} (1979) 57.

\bibitem{BBG} 
W.A. Bardeen, A.J. Buras and J.-M. G\'erard,
Phys.\ Lett.\ B {\bf 180} (1986) 133,
Nucl. Phys. {\bf B 293} (1987) 787; 
Phys. Lett. {\bf B 192} (1987) 138; 
Phys. Lett. {\bf B 211} (1988) 343;
A.J. Buras and J.-M. G\'erard, Nucl. Phys. B (Proc. Suppl.)
{\bf 7A} (1989) 375; 
J.-M. G\'erard, Acta Phys. Polon. {\bf B 21} (1990) 257;
A.J. Buras in ``CP Violation'', p. 575, C. Jarslkog (ed),
World Scientific (1989).

\bibitem{BAR89} 
W.A. Bardeen,
Nucl. Phys. B (Proc. Suppl.) {\bf 7A} (1989) 149.

\bibitem{BAR99} 
W.A. Bardeen in ``Kaon Physics'', p. 171,   
J.L. Rosner and B.D. Winstein (eds), Univ. Chicago Press  (2001).

\bibitem{BP95}
J. Bijnens and J. Prades, Phys. Lett.
{\bf B 342} (1995) 331;
Nucl. Phys. {\bf B 444} (1995) 523.

\bibitem{BP00} 
J. Bijnens and J. Prades,
J. High Energy Phys. {\bf 01} (2000) 002.

\bibitem{ENJL} 
J. Bijnens, C. Bruno and E. de Rafael,
Nucl. Phys. {\bf B 390} (1993) 501;
J. Prades, Z. Phys. {\bf C 63} (1994) 491
[Erratum Eur. J. Phys. {\bf C 11} (1999) 571];
J. Bijnens, E. de Rafael and H.q. Zheng,
Z. Phys. {\bf C 62} (1994) 437;
J. Bijnens and J. Prades, Phys. Lett.
{\bf B  320} (1994) 130;
Z. Phys. {\bf C 64} (1994) 475;
Nucl. Phys. B (Proc. Suppl.) {\bf 39BC}
(1995) 245;
J. Bijnens,  Phys. Rept.
{\bf 265} (1996) 369.

\bibitem{PR00} 
S. Peris and E. de Rafael,
Phys. Lett. {\bf B 490} (2000) 213
[Erratum, hep-ph/0006146]. 

\bibitem{CP03} 
O. Cat\`a and S. Peris,
J. High Energy Phys.  {\bf 03} (2003) 060.

\bibitem{BGLP03} 
J. Bijnens, E. G\'amiz, E. Lipartia  and J. Prades, 
J. High Energy Phys. {\bf 04} (2003) 055.

\bibitem{BGP06a} 
J. Bijnens, E. G\'amiz and J. Prades,
in preparation.

\bibitem{Q7Q8}
J. Bijnens, E. G\'amiz and J. Prades,
J. High Energy Phys.  {\bf 10} (2001) 009.


\bibitem{BP03} 
J. Bijnens and J. Prades, presented at
the second EURIDICE collaboration meeting, 6-8 february 2003, Orsay.

\bibitem{PBG05} 
J. Prades, J. Bijnens and E. G\'amiz,
``Trento 2004, Large $N_c$ QCD'',  p. 179-190,
J.L. Goity {\em et al} (eds.),
World Scientific (2005),  hep-ph/0501177. 

\bibitem{NLO} 
A.J. Buras, M. Jamin and P.H. Weisz,
Nucl.\ Phys.\ B {\bf 347} (1990) 491;
S. Herrlich and U. Nierste,
Nucl.\ Phys.\ B {\bf 419} (1994) 292; 
Phys.\ Rev.\ D {\bf 52} (1995) 6505; 
Nucl.\ Phys.\ B {\bf 476} (1996) 27;
M. Ciuchini, E. Franco, V. Lubicz, G. Martinelli, 
I. Scimemi and L. Silvestrini,
Nucl.\ Phys.\ B {\bf 523} (1998) 501.

\bibitem{CDG00} 
V. Cirigliano, J.F. Donoghue and E. Golowich,
J. High Energy Phys. {\bf 10} (2000) 048.

\bibitem{BGP06b} 
J. Bijnens, E. G\'amiz and J. Prades,
in preparation.

\bibitem{MOU95} 
B. Moussallam and J. Stern, 
hep-ph/9404353;
B. Moussallam, Phys. Rev. {\bf D 51}
(1995) 4939;
Nucl. Phys. {\bf B 504} (1997) 381.

\bibitem{KN01} 
M. Knecht and A. Nyffeler,
Eur. Phys. J. {\bf C 21} (2001) 659.

\bibitem{CIR04}
V. Cirigliano, G. Ecker, M. Eidem\"uller, A. Pich and J. Portol\'es,
Phys. Lett. {\bf B 596} (2004) 96.

\bibitem{RPP03}
P.D. Ruiz-Femen\'{\i}a, A. Pich and J. Portol\'es,
J. High Energy Phys.  {\bf 07} (2003) 003;
Nucl.  Phys. B (Proc. Suppl.)  {\bf 133} (2004) 215.

\bibitem{ABT01} 
G. Amor\'os, J. Bijnens and P. Talavera,
Nucl. Phys.  {\bf B 602} (2001) 87.

\bibitem{BT02} 
J. Bijnens and P. Talavera, 
J. High Energy Phys. {\bf 03} (2002) 046;
Nucl. Phys. {\bf B 489} (1997) 387.

\bibitem{CENP03} 
V. Cirigliano, G. Ecker, H. Neufeld and A. Pich,
J. High Energy Phys.  {\bf 06} (2003) 012.

\bibitem{staggered}
W. Lee et al., Phys. Rev. {\bf D 71} (2005) 094501.

\bibitem{AOK04}
Y. Aoki {\it et al.},
Phys.\ Rev.\ D {\bf 72} (2005) 114505.

\bibitem{AOK05} 
Y. Aoki et al., 
hep-lat/0508011.

\bibitem{latticeunquenched}
E.~G\'amiz, S.~Collins, C.~T.~H.~Davies,
J.~Shigemitsu and M.~Wingate  [HPQCD Collaboration],
PoS {\bf LAT2005} (2005) 47;    
T.~Bae, J.~Kim and W.~Lee,
\emph{ibid} 335; 
\emph{ibid} 338; 
S.~Cohen, \emph{ibid} 346;
F.~Mescia, V.~Gim\'enez, V.~Lubicz, 
G.~Martinelli, S.~Simula and C.~Tarantino, 
\emph{ibid} 365;
J.~M.~Flynn, F.~Mescia and A.~S.~B.~Tariq  [UKQCD Collaboration],
J. High Energy Phys. {\bf 11} (2004) 049. 

\end{thebibliography}
\end{document}